\documentclass[aps,twocolumn,groupedaddress]{revtex4} 

\usepackage{graphicx}  
\usepackage{dcolumn}   
\usepackage{bm}        
\usepackage{amssymb, amsmath}   

\usepackage{setspace}
\usepackage{natbib}

\usepackage{float}
\usepackage[caption=false]{subfig}

\begin{document}


\hspace{5.2in} \mbox{}

\title{Single molecule NMR detection and spectroscopy using single spins in diamond}

\author{V. S. Perunicic}
\email{vpe@unimelb.edu.au}
\author{L. T. Hall}
\author{D. A. Simpson}
\author{C. D. Hill}
\author{L. C. L. Hollenberg}
\affiliation{Centre for Quantum Computation and Communication Technology, School of Physics, University of Melbourne, Victoria 3010, Australia }

\date{\today}

\begin{abstract}
Nanomagnetometry using the nitrogen-vacancy (NV) centre in diamond has attracted a great deal of interest because of the combined features of room temperature operation, nanoscale resolution and high sensitivity. One of the important goals for nano-magnetometry is to be able to detect nanoscale nuclear magnetic resonance (NMR) in individual molecules. Our theoretical analysis shows how a single molecule at the surface of diamond, with characteristic NMR frequencies, can be detected using a proximate NV centre on a time scale of order seconds with nanometer precision.  We perform spatio-temporal resolution optimisation and also outline paths to greater sensitivity. In addition, the method is suitable for application in low and relatively inhomogeneous background magnetic fields in contrast to both conventional liquid and solid state NMR spectroscopy.
\end{abstract}

\pacs{}
\maketitle
\linespread{0.9}

Magnetic resonance (MR) based detection and imaging is an important tool across many areas of nanoscience. From a bio-medical perspective, the need to better understand cellular processes at the nanoscale, occurring both naturally and as a result of introduced nanoparticles and/or molecular species, poses a significant and a constant question. The long tradition of magnetometry techniques in bio-imaging, such as in electron spin resonance (ESR), nuclear magnetic resonance (NMR) and magnetic resonance imaging (MRI), have been successful in detecting the bulk properties of cells and their reactions \cite{Borbat2001, Mittermaier2006}. However, all of these methods rely on detecting very large numbers of electronic or nuclear spins and hence are limited fundamentally in their resolution. New methods such as magnetic resonance force microscopy (MRFM) are capable of detecting single electron or nuclear spins, yet have the additional requirements of vacuum conditions and low temperatures $(<2K)$ \cite{RugarMRFM04,Vinante11,Mamin05}.

\begin{figure}[t]%
\centering
 \includegraphics[width=1\linewidth]{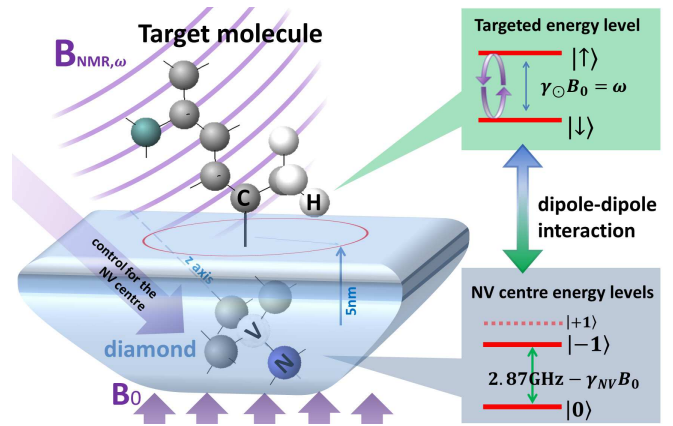}
 \vspace{-15pt}
  \caption{Schematic representation of the system containing an NV centre and a single target molecule. The NV centre is located in a diamond lattice.  The microwave control is set to address only two of the centre's three energy states; the NV centre exhibits a dipole-dipole interaction with a target's nuclear spin. While the target is being subjected to an oscillating magnetic field $B_{\rm NMR}$ resonant with its (known) energy level separation. }\label{fig:1}
   \vspace{-15pt}
\end{figure}

We focus on how to perform NMR for the task of detecting individual molecules under ambient conditions. The ability to selectively detect molecular species has important implications in a range of fields, including nanomedicine. The MR detector we consider is the nitrogen-vacancy (NV) centre in diamond, in which a number of fortuitous properties converge making it a very promising sensor: it is bio-compatible, exhibits sustained fluorescence over arbitrarily long timescales, and is inherently a nanoscale magnetic sensor with high sensitivity. Recent experiments demonstrate that this centre in diamond is capable of identifying the presence of relatively modest number of nuclear spins (between $10^4$ and $10^6$ actual protons) external to the diamond lattice under controlled conditions, both by passive observation \cite{Staudacher13} and by manipulation of proton spins states \cite{Mamin13}.
The NV centre is a spin 1 system which ground state can be optically initialised and read out \cite{Wrachtrup93,Gruber97,Jelezko04,Harrison04,Manson06}. The coherence of a centre several nanometres below a high purity diamond surface is relatively long at room temperatures extending into millisecond timescale \cite{Balasubramanian09}. The NV centres are used as florescent bio-markers \cite{Chang08,Neugart07}, which spin state can be coherently controlled inside living cells \cite{McGuinness11}.
Their applications span numerous fields including quantum computing, encryption \cite{Dutt07,Ladd10,Beveratos02} and single photon sources \cite{Kurtsiefer00}. Developments in the field of magnetometry can broadly be described in terms of DC and AC field sensing \cite{Balasubramanian08,Taylor11} using single NV centres or ensembles \cite{Steinert10}. For random fields, particularly those encountered in biology, schemes for decoherence based magnetometry have been proposed \cite{Cole09,HallPRL09,Hall10IonCh,Meriles10}. Interest in different types of electronic spin detection using NV centres is rapidly growing \cite{Grotz11,McGuinness13NJP,MaminEspin12,Grinolds13,Steinert12}, while in the biological context, the detection of atomic spin labels in an artificial cell membrane using the NV centre as a probe has recently been demonstrated \cite{Kaufmann13}. Recent measurement have detected ferritin molecules based on their carriage of a large number of electronic spins \cite{ErmakovaFerritin13}, an important goal is to be able to use the NV centre as a nanoscale NMR magnetometer for the task of selective single molecule detection. At present, methods to detect nonspecific single and paired nuclear spins within the diamond lattice are being developed \cite{Zhao12Arxiv,Kolkowitz12Arxiv}.
Recently, methods have been proposed for external molecule detection using the NV centre based on detecting the natural flip-flop frequency of nuclear dimer spins \cite{Zhao2011} or matching the Rabi frequency of the NV centre with the energy levels of a target spin \cite{Cai11Arxiv}. In the approach we consider the external target spins in the molecular system are driven and from the NV signal we determine the anisotropic NMR spectrum, which then provides selective molecular detection.

The overall method is depicted schematically in Figure 1. It involves continuously subjecting the particular target molecular species to an NMR field, while simultaneously performing a particular control sequence (e.g. spin-echo, CPMG, Uhrig \cite{Hahn50,Meiboom58,Uhrig07}) on the NV centre. Application of the NMR field (driving field) of desired strength $B_{\rm NMR}$, which is resonant with a particular transition characteristic to the target molecule, ensures that the detection is sensitive to that molecular species only. When on resonance the targeted nuclear spin(s) undergoes Rabi oscillations, thus producing an fluctuating magnetic field at the NV centre, as represented in Fig. 2a. This fluctuating field is subsequently sensed through a measurement sequence applied to the NV spin, which converts the accumulated phase to a state population difference. The population of the NV spin states $S$ can be optically measured, and the change in the population $\Delta S$ is defined as the signal. Continuous target driving can additionally provide a simple way of decoupling the targeted nuclei from the environment thus increasing their coherence times.

\begin{figure}[t]%
\centering
  \includegraphics[width=1\linewidth]{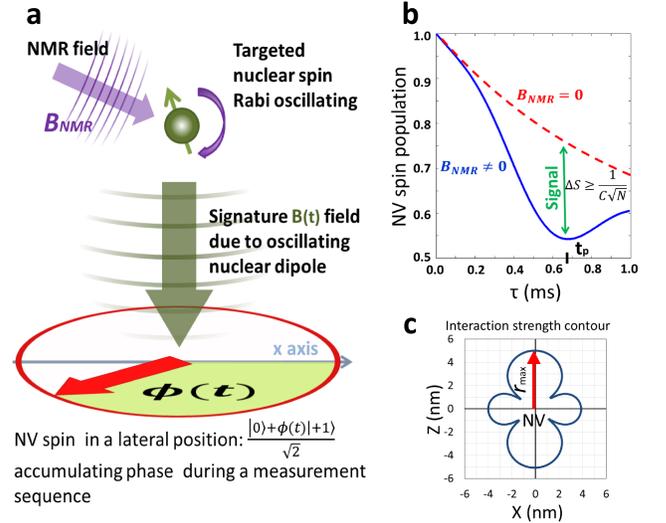}
   \vspace{-25pt}
  \caption{Measurement protocol. a) The resonant driving field $B_{NMR}$ induces oscillations in the specific targeted nuclear spin transition. The dipole of a targeted nucleus creates an fluctuating field $B(t)$, inducing a phase $\phi(t)$ in the NV spin which is transformed into change in the spin state population $\Delta S$ (signal). b) The spin population of the NV centre under a spin-echo sequence in the proximity (5nm) of a single molecule proton spin with (dashed, red) and without (solid, blue) the driving field. The signal is nonzero when the driving field is resonant to a targeted NMR transition. c) The cross-section of the interaction strength $k(\mathbf{r})$ contour between the targeted nuclear spin and the NV centre. The maximum distance between the NV and the target $r_{\rm max}$ is defined as an intersection of the contour and the z-axis.}\label{fig:2}
 \vspace{-15pt}
\end{figure}

Initially, the NV probe can be characterised when no driving field is present ($B_{\rm NMR}=0$), reflecting the influences of the entire environment that the probe is exposed to. The system can then be analysed whilst the target molecules subjected to NMR ($B_{\rm NMR}\neq0$), as illustrated in Fig. 2b. By comparing the outcomes of the two cases information about the target molecule is extracted.

To analyse the technique we consider the following target species as proof of principle examples: aldehyde $\rm{(-CHO)}$, hydroxymethyl $\rm{(-CH_2OH)}$, and methyl $\rm{(-CH_3)}$ groups attached to the diamond surface. The aldehyde group showcases the detection of only one external proton spin, while the other two examples represent detection of two and three proximate protons. The NV centre is positioned several nanometres below an individual target as depicted in Figure 3. We are primarily interested in how to obtain a NMR spectrum of each individual group and how that information can be used to quickly detect the presence of a targeted molecule attached to the diamond surface.
The Hamiltonian of the combined system is given by:

\begin{align} \label{eq:1,2,3,4}
&H_{\rm{system}}=H_{\rm nv}+H_{\odot}+H_{d-d},\nonumber \\
&H_{\rm nv}=\hbar\gamma_{\rm nv}B_0Z_{\rm nv} +\Delta Z_{\rm nv}^2 +\Pi(t),\nonumber \\
&H_{\rm \odot}=\sum_{i} \hbar\gamma_{\rm \odot}^i( B_0Z^i_{\rm \odot}+ B_{\rm NMR}[X^i_{\rm \odot}\cos(\omega t)
+ Y^i_{\odot}\sin(\omega t)]),\nonumber \\
&H_{d\mbox{-}d}=\sum_{i}\gamma_{\rm nv}\gamma_{\odot}^i \hbar^2 \frac{\mu_0}{4\pi}\frac{1}{r^3}[\mathbf{S_{nv}\cdot S_{\odot}^{\it{i}}}
-\frac{3}{r^3}(\mathbf{S_{nv}\cdot r})(\mathbf{S_{\odot}^{\it{i}}\cdot r})] \nonumber \\
&\quad \quad \quad +\sum_{i,j}\gamma_{\odot}^i\gamma_{\odot}^j \hbar^2 \frac{\mu_0}{4\pi}\frac{1}{r^3}[\mathbf{S_{\odot}^{\it{i}}\cdot S_{\odot}^{\it{j}}}
-\frac{3}{r^3}(\mathbf{S_{\odot}^{\it{i}}\cdot r})(\mathbf{S_{\odot}^{\it{j}}\cdot r})],
\end{align}
where $H_{\rm nv}$ encompasses the NV centre's zero field splitting $\Delta=2.87\rm GHz$, it's interaction with the background $B_{0}$ and NV-control fields $ \Pi(t)$; $H_{\odot}$ is the interaction of each target nuclear spin $i$ with the background field and the driving field of frequency $\omega$; and $H_{d-d}$ is the total dipole interaction between the NV spin and each of the target nuclear spins as well as the interaction between target nuclei themselves. In the subsequent analysis the system based on this Hamiltonian will be solved numerically using a master equation to describe decoherence effects on the target and NV spins. However, it is instructive to analytically obtain, using the secular approximation, the interaction strength $k(\mathbf{r})$ between a NV centre and one nuclear spin:
\begin{equation}\label{eq:6}
\setlength{\abovedisplayskip}{5pt}
k(\mathbf{r})=\gamma_{\rm nv}\gamma_{\odot}\hbar \frac{\mu_0}{4\pi}\frac{1}{r^3}\left( 1-\frac{3r_z^2}{r^2}\right).
\end{equation}
Figure 2c  shows a lobe-like contour of the function $k(\bm{r})$. As the shells of equivalent interaction strength are not radially symmetric, it is convenient to define the distance between the NV centre and the target $r_{\rm max}$ as the z-axis intercept of the $k(\bm{r})$  shell on which the target is located, as depicted in Fig. 2c.

Traditionally, NMR techniques can be classified by the environment as either liquid or solid state. These two cases are fundamentally different from the perspective of the target NMR profile \cite{Schroder04,Ingo05,Cremer07} and require further insight in light of using the NV centre as a detection probe.

A liquid NMR spectrum is dominated by isotropic J-coupling and chemical shift while any other anisotropic interactions including dipole-dipole (d-d) coupling typically average out due to constant Brownian rotation with respect to the background magnetic field. Thus, molecules in a solution possess a relatively simple spectrum with sharp resonant peaks particularly at high background fields. Consequently, the coherence times of the proton spins in liquids tend to be very long. Being highly localised a NV centre would in this case require a high concentration of target molecules for a successful detection process.

A solid state NMR spectrum is strongly dominated by a d-d interaction while the influence of J-coupling and chemical shift depend on the molecule and the background magnetic field. It is the dominant nature of the anisotropic interactions that makes the solid state NMR spectrum relatively broadened, yet rich with spatial information. In a powdered form, an ensemble of possible molecular orientations creates wide regions of resonant frequencies instead of single peaks. Single crystals have more defined peaks dependent on orientation, nonetheless they are not a suitable bio-molecular platform. A lot of effort has been dedicated to eliminating anisotropic coupling in non-crystalline samples, customarily requiring bulky equipment, and sophisticated techniques \cite{Ingo05}.

The molecules bonded to the surface of diamond are effectively stationary and in the following circumstances do not exhibit anisotropic broadening. If the background magnetic field is constant over a diamond surface, then the orientation of each surface bonded molecule with respect to the background field will be the same. These requirements are not difficult to achieve locally since the detection area, directly above the NV centre is of a nanometre scale. In addition, when the surface density of the molecules is not large (e.g. $<1/\mathrm{nm}^{2}$), the nuclear coherence is relatively long ($> 1\rm ms$). Additional information about the diamond surface chemistry and its potential for a bio-chemical applications can be found in reference \cite{Krueger2012}. However, the number of molecules bonded to the microscopic diamond surface is too small for conventional NMR characterisation. Evidently in this work we need a different approach in obtaining NMR spectrum of a target species. A NV centre experiencing a weak d-d interaction with a target molecule will be used as a quantum probe.

\begin{figure}[h]%
\centering
  \includegraphics[width=1\linewidth]{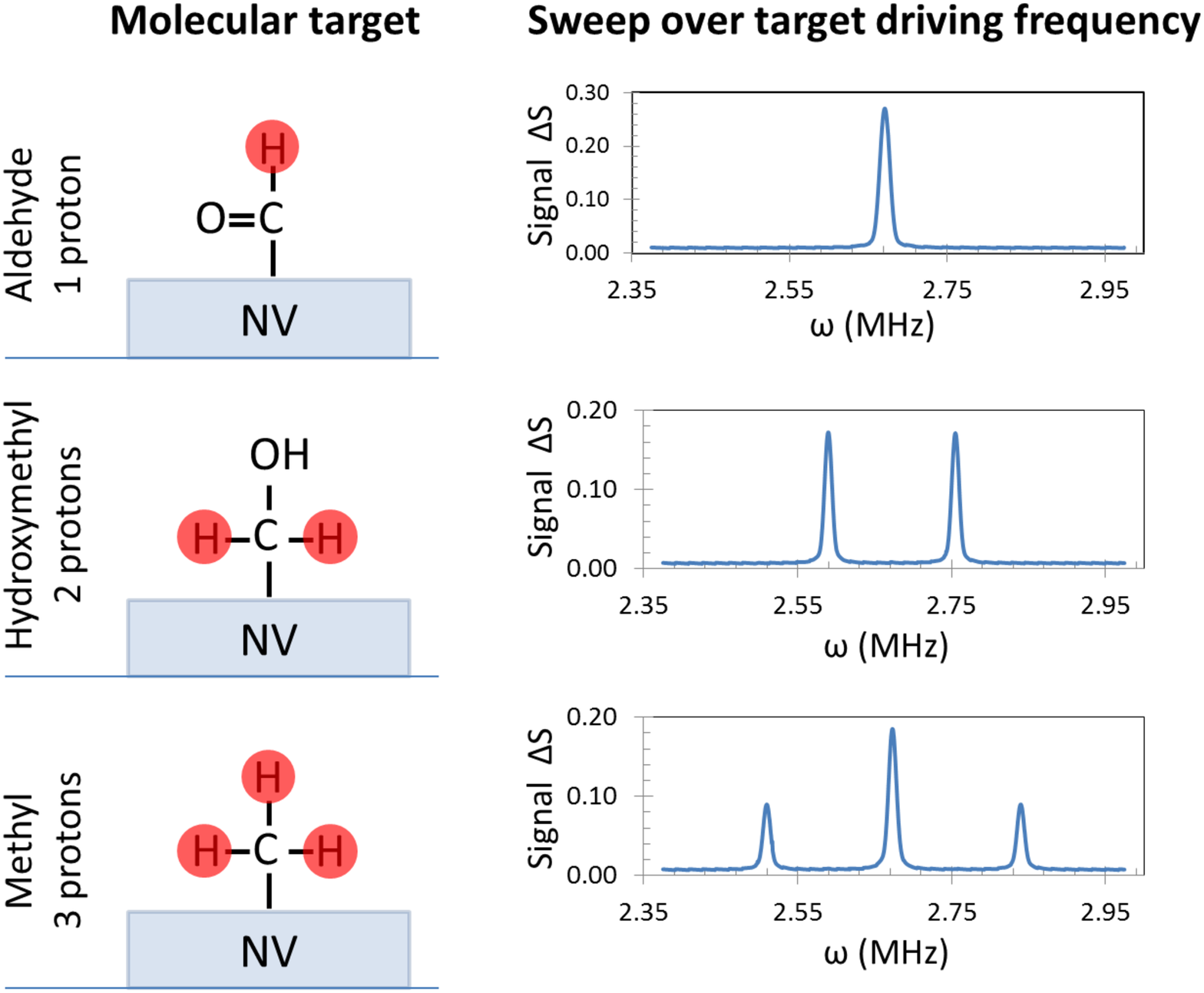}
   \vspace{-20pt}
  \caption{Measurement spectrum of three single species: aldehyde, hydroxymethyl, and methyl groups, obtained using the NV centre. The groups are attached to the surface of the diamond 5nm above the NV centre.  The spectra of the highlighted protons is depicted on the graphs. The hydroxyl proton of the hydroxymethyl group is inactive depicting a consequence of proton exchange in a aqueous solution. Equivalently, the rest of the diamond surface can also be terminated in with NMR neutral groups (e.g. carboxyl). Note, the background magnetic field is directed along the molecule-diamond covalent bond.}\label{fig:3}
   \vspace{-7pt}
\end{figure}

The target molecule NMR spectrum characterisation is as follows. A NV centre is characterised to a high precision without a driving field present using a preferred control sequence. We demonstrate the method using spin-echo, however higher order CPMG or Uhrig sequences can increase sensitivity by virtue of extending the coherence time of the NV centre \cite{Taylor08,Hall10,Naydenov11,deLange11}. Assuming the background magnetic field is aligned with the NV centre's quantisation axis the spin population will simply reflect the decoherence envelope. Thus, it is sufficient to choose a single sensitive time point $t_{\rm p}$ on the envelope as illustrated in Fig. 2.b. Then, a sweep over a driving frequency $\omega$ is performed while the NV centre is simultaneously observed at the point $t_{\rm p}$  as per the chosen control sequence. When the driving field becomes resonant with a particular nuclear transition the signal from the NV centre will increase corresponding to a peak in the NMR spectrum.

Figure 3 shows the numerically simulated NMR spectra of a single chemical group characterised using a NV centre located 5nm below the surface of an ultra pure bulk diamond. The spin echo sequence was applied to the NV centre while the resonant field was continuous. The control pulses for the NV centre were assumed to be perfect. A full dipole-dipole interaction between all the nuclear spins was considered, while decoherence for both the NV centre and the target nuclei was introduced through Lindblad terms \cite{BreuerPetrucci}. Initial conditions of the target nuclei was an equally mixed state, reflecting a high temperature environment. It is worth noting that this analysis was done for a low background field, $B_{0}=0.01\mathrm{T}$, where d-d interaction between the target species nuclei is the most dominant in comparison to other inter-nuclear interactions. The results demonstrate a sufficient level of sensitivity to perform a single molecule NMR spectroscopy, while multiple targets contribute to the signal strength in a simple manner. However, when using the higher order sequences (CPMG or Uhrig) to further improve the sensitivity, a trade-off may need to be reached, as the Rabi frequency of the target spin has to be synchronised with the control pulses of the NV centre sequence \cite{Taylor11,HallPRL09,Hanson06}. For higher order sequences this requires a stronger driving field which is in turn a limiting factor either because of broadening or experimental considerations.

Once the NMR spectrum of the target molecules is characterised, the NV centre in diamond can be used as a quick and selective single molecule detector. The proposed method is as follows. A number of resonant frequencies that identify particular species are selected. Then, the signal is screened for only those particular frequencies. If the targeted species is present the signal is observed. There are a number of approaches to analysing the change in population depending on the shape of the signal when no driving field is present.  In line with the previous description it is sufficient to consider a single temporal point $t_{\rm p}$. For each individual targeted transition a resonant field of strength $B_{\rm NMR}$ is applied while the NV spin population is observed for a control sequence of length $t_{\rm p}$. Detection of a target species with this particular transition is achieved if the signal $\Delta S$ is observed. As the signal is detected optically, its uncertainty is shot noise limited. Conformation of the presence of the target can be defined as the moment when the signal becomes resolvable with respect to its measurement error:
\begin{eqnarray} \label{eq:5}
\Delta S=S(B_{\rm NMR}=0)-S(B_{\rm NMR}\neq0)\geq\frac{1}{C\sqrt{N}},
\end{eqnarray}
where $N$ is the number of measurements and $C$ is related to the photon collection efficiency \cite{Taylor11}. The outlined method also provides a temporal resolution as the total time it takes to detect a targeted transition becomes $T=Nt_{\rm p}$. The magnitude of the signal depends on the interaction strength $k(\bm{r})$ between the target nucleus and the  NV centre, which is distance dependent, providing the spatial resolution of the method.

\begin{figure}[h]%
\centering
  \includegraphics[width=1\linewidth]{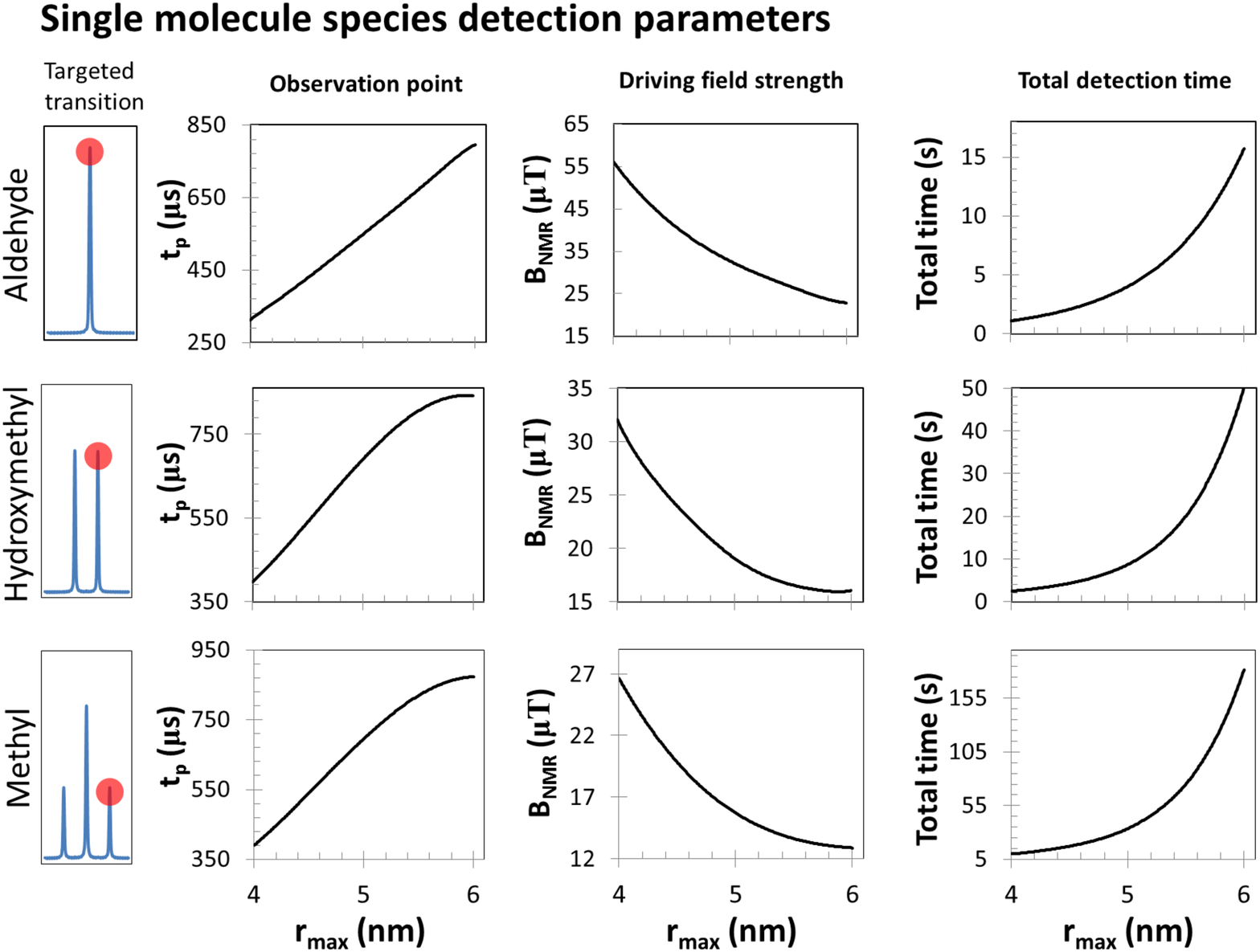}
   \vspace{-20pt}
  \caption{Optimised detection time as a function of distance for a highlighted proton transition demonstrated that the second timescale is achievable using a robust version of the detection method: continually driven targeted species and a spin-echo measurement sequence for the NV centre.  The optimal parameters for such detection are also shown corresponding to the driving field strength $B_{NMR}$ and the time point $t_{\rm p}$ at which the NV centre's signal is monitored.}\label{fig:4}
\end{figure}

The signal strength is dependent on many factors: the target distance $r_{\rm max}$, the time point $t_{\rm p}$, the strength of the driving field $B_{\rm NMR}$, the coherence times of the NV centre/target species and the number of target molecules in the vicinity of the NV centre. It also depends on the type of the targeted transition and the number of nuclei it involves. To explore the full potential of the detection method we investigate the relationship between these parameters. Figure 4 shows the synthesis of the system's complex parameter space. Three different transitions have been considered under a spin-echo sequence for the NV centre. For each, the relationship between $r_{\rm max}$, $t_p$ and $B_{\rm NMR}$ was analysed including an optimization for the temporal resolution. The parameters chosen for the simulations relating both to Figures 3 and 4 reflect the presently achievable experimental conditions as follows: $B_0=0.01\mathrm{T}$, $C=0.05$, $T_{\rm 1NV}=5\mathrm{ms}$, $T_{\rm 2NV}=1\mathrm{ms}$ \cite{Naydenov11,Maze08,Balasubramanian08,Balasubramanian09,Childress06}, $T_{1 \odot}=10\mathrm{ms}$, $T_{2\odot}=1\mathrm{ms}$. A significant limiting factor to the present NV-based detection is the centre's transverse coherence time $T_{\rm 2NV}$, which for the shallow NV centres is strongly influenced by the diamond surface conditions. Figure 5 shows the trend of the total detection time associated with decrease in $T_{\rm 2NV}$ for a single proton target over a rage of distances. These results indicate that the detection method may be used under a range of accessible coherence conditions, however, there is a decrease in temporal resolution with the loss of coherence.

\begin{figure}[h]%
\centering
  \includegraphics[width=0.8\linewidth]{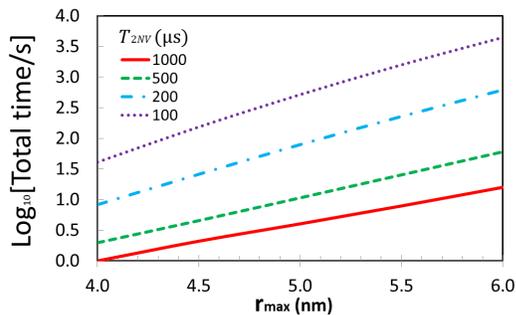}
   \vspace{-10pt}
  \caption{Optimised detection time for sensing a single proton over a range of distances ($r_{\it max}$) from the NV centre for several transverse coherence times of the NV centre.}\label{fig:4}
\end{figure}

There is a large scope for the application of single molecule detection using the NV centre. Consider detecting molecules in process of bonding on the diamond surface by screening over multiple signature frequencies, or monitoring just one NMR transition of an intermediate (antenna) molecule which depends on a particular external factor. The antenna molecule may bond with a target molecule or change its tertiary structure.
Simultaneous observation of more than one transition within a single molecule increase the signal accordingly. In addition to parameter optimisation and the use of higher order control sequence on the NV centre, the detection sensitivity can be further increased by introduction of NMR pulses onto the target spin. This is a less robust approach relying on a precise target spin control. We note, that additional improvements to sensitivity may naturally come from making the experimental parameters (e.g. photon collection efficiency, spin coherence) more favourable.

In conclusion, we have investigated the application of NV sensing to single molecule NMR through direct and selective driving of the molecular nuclear transitions. Our results shows that single nuclear spin detection in selected molecular targets is possible with temporal resolution in a range of seconds.

This work was supported by the Australian Research Council under the Centre of Excellence scheme (project number CE110001027).


\end{document}